\documentclass[preprintnumbers]{revtex4}
\usepackage{color}
\usepackage{graphicx}
\usepackage{eurosym}
\usepackage{graphicx}
\usepackage{amsmath,amssymb}
\usepackage{color}
\usepackage[outdir=./]{epstopdf}
\usepackage{amsmath}
\usepackage{amsfonts}
\usepackage{amssymb}
\usepackage{eurosym}
\usepackage[utf8]{inputenc}
\usepackage{amsmath,amssymb}
\usepackage{graphicx}
\usepackage{amsmath}
\usepackage{xcolor}
\usepackage{graphicx}
\usepackage{hyperref}

\graphicspath{ {./images/} }

\DeclareSymbolFont{matha}{OML}{txmi}{m}{it}
\DeclareMathSymbol{\varv}{\mathord}{matha}{118}
\graphicspath{ {./images/} }
\begin{document}

\date{\today }
\title{Bound state solutions of the two--dimensional Schr\"{o}dinger equation
with Kratzer--type potentials}
\author{Roman Ya. Kezerashvili$^{1,2,3}$, Jianning Luo$^{1}$, and Claudio R.
Malvino$^{1}$}
\affiliation{$^{1}$New York City College of Technology, The City University of New York,
Brooklyn, USA\\
$^{2}$The Graduate School and University Center, The City University of New
York, New York, USA\\
$^{3}$Long Island University, New York, USA}

\begin{abstract}
Exactly solvable models play an extremely important role in many fields of
quantum physics. In this study, the Schr\"{o}dinger equation is applied for
a solution of a two--dimensional (2D) problem for two particles interacting
via Kratzer, and modified Kratzer potentials. We found the exact bound state
solutions of the two--dimensional Schr\"{o}dinger equation with Kratzer--type
potentials and present analytical expressions for the eigenvalues and
eigenfunctions. The eigenfunctions are given in terms of the associated
Laguerre polynomials.
\end{abstract}

\maketitle

\section{Introduction}

The investigation and understanding of a quantum mechanical system is
dependent on garnering solutions to the Schr\"{o}dinger equation. Solutions
to the equation provides the eigenvalues and eigenfunctions of the system,
which defines the fundamental information needed to understand the system.
There are only several potentials for which there are explicit solutions for
all partial waves or an exact solution of the $s-$wave Schr\"{o}dinger
equation in three-dimensional (3D) space \cite%
{Landau,Schiff2014,Davydov1975,Sakurai1985,Fluge,Robinett2006,Griffiths2018,Suslov2023}%
.

As scientific advances continue to be made in the fields of transistors,
material sciences, and semiconductors, it has become increasingly valuable
to conduct quantum mechanics in two--dimensional space \cite%
{Gugiuzza2017,Wachter2017,Robinett2006}. The most prominent field of study
for the use of 2D quantum mechanics are monolayer materials. Two decades
since the Noble prize winning discovery of monocrystalline graphitic films,
by Novoselov and Geim in 2004 \cite{Novoselov2004}, the study of
two--dimensional monolayer materials have expanded far beyond monoatomic
materials, such as graphene.

Although transition metal dichalcogenides have been known and studied in
bulk form since the 1960's \cite{Frindt1966}, it was not until 2010, that
the full spectrum of applications for 2D monolayer transition metal
dichalcogenides were realized \cite{Mak2010}. It was then that the
application of transitional metal dichalcogenides, such as WSe$_{2}$, WS$%
_{2} $, MoSe$_{2}$, MoS$_{2}$, TeSe$_{2}$, and TeS$_{2}$, as semiconductors
became established due to breakthroughs in understanding their properties at
the monolayer scale \citep{Dai2016,Avouris2017}. These molecular layered 2D
systems received considerable attention %
\citep{Mak2010,Cheng2014,Lee2014,Cotlet2016,Manzeli2017,Berman2016,Berman2017,Wang2018,Brunetti2018,RKAS2021b}%
. We cite these works, but the recent literature on the subject is not
limited by them. Most of the research was focused on optical phenomena,
Bose-Einstein condensation of exciton, magnetic properties of excitons, et
al. One can consider the application of the Kratzer--type potentials for
descriptions of vibrational and rotational energies for the two--dimensional
molecules of WSe$_{2}$, WS$_{2}$, MoSe$_{2}$, MoS$_{2}$, TeSe$_{2}$, and TeS$%
_{2}$ in 2D semiconductors. The Kratzer \cite{Kratzer1926} and modified
Kratzer--type\ \cite{LeRoy1970,Berkdemir2006} potentials is mostly applied in
atomic physics, molecular physics, and quantum chemistry. They are used to
describe the interactions of molecular structure in 3D quantum mechanics.
These potential are reliable in terms of obtaining eigenvalues and
eigenfunctions for vibrational and rotational energies. The Kratzer
potential is known to approach infinity when the internuclear distance in
molecules approaches zero, due to the repulsion that exist between the
molecules. As the internuclear molecular distance approaches infinity, the
potential approaches to zero. In Ref. \cite{Molas2019} was suggested the
modified Kratzer potentional for description of energy levels of the Rydberg
2D excitons.

In Ref. \cite{Suslov2023} using the Nikiforov-Uvarov (NU) method \cite%
{NikUvarov} are given the summary for analytical solutions of Schr\"{o}%
dinger equations in 3D space for $s-$ and arbitrary $l$ states for basic
potentials employed in non--relativistic quantum mechanics. Solutions of
two--dimensional Schr\"{o}dinger equation for two particles are obtained in
the close form \cite{Kezerashvili2023} within the framework the NU paradigm 
\cite{NikUvarov}. The decrease of the dimensionality from 3D to 2D decreases
the degree of freedom by one and, hence, the kinetic energy of the
particles. Thus, the reduction of dimensionality decreases the kinetic
energy of two particles in 2D configuration space due to the decrease of the
degrees of freedom from three to two \ \cite%
{kezFB2019,Kezerashvili2020FewBody}. The latter leads to the change of the
energy spectrum for two interacting molecules in 2D space. Using the
approach \cite{Fluge} applied in 3D space, below we are finding the closed
solution for the 2D Schr\"{o}dinger equation for the Kratzer and modified
Kratzer potentials.

\section{Kratzer and Kratzer--type potentials}

To study the rotation-vibration spectrum of a diatomic molecule Kratzer in
1920 introduced the following central symmetry potential \cite{Kratzer1926} 
\begin{equation}
V(r)=-2D_{0}\left( \frac{r_{0}}{r}\ -\frac{1}{2}\ \frac{r_{0}^{2}}{r^{2}}\
\right) ,  \label{Kr1}
\end{equation}%
where $D_{0}$ is the chemical dissociation energy of the lowest vibrational
level which determines the depth of the potential and the parameter $r_{0}>0$
is the equilibrium internuclear separation. $D_{0}$ differs slightly from
the electronic (or spectroscopic) dissociation energy of the diatomic
molecule $D_{e}$ used in some works with potential (\ref{Kr1}). This
potential has a minimum $V(r_{0})=-D_{0}$ and contains both a repulsive part
and a long-range attraction. For the Kratzer potential, as $r\rightarrow 0$, 
$V(r)\rightarrow \infty $ due to the internuclear repulsion and as $%
r\rightarrow \infty $, $V(r)\rightarrow 0$ i.e., the molecule decomposes.

To study the rotation-vibration spectrum of diatomic molecules the central
symmetry Kratzer potential \cite{Kratzer1926} was modified by a long-range
interatomic potential \cite{Berkdemir2006} that has wide applications in
chemical physics 
\begin{equation}
V(r)=D_{0}\left( \frac{r-r_{0}}{r}\right) ^{2}.  \label{Kr31}
\end{equation}%
This potential presents the modified Kratzer potential and is shifted in
amount of $D_{0}$. In Refs. \cite{Berkdemir2006,Doma2023} two-particle
problem with the modified Kratzer potential (\ref{Kr31}) is solved in 3D
space using the NU method.

Improvements in semiconductor growth techniques over the subsequent decades,
which enabled the manufacture of effectively two--dimensional structures, led
to a resurgence of excitons in 2D materials. An exciton is the bound state
of an electron and hole in bulk, two--dimensional and even in one--dimensional
materials. The Schr\"{o}dinger equation were applied to problems of excitons
in two dimensions.\ It was suggested the modified Kratzer potential \cite%
{Molas2019} for a formation of the Rydberg excitons in 2D materials 
\begin{equation}
V(r)=-\mathfrak{K}\frac{e^{2}}{\rho }\left( \frac{r_{0}}{r}\ -g^{2}\ \frac{%
r_{0}^{2}}{r^{2}}\ \right) .  \label{Kr25}
\end{equation}%
In Eq. (\ref{Kr25}) $\mathfrak{K}=\frac{1}{4\pi \varepsilon _{0}}=9\times
10^{9}Nm^{2}/C^{2}$ with the dielectric permittivity of vacuum $\varepsilon
_{0}$, $\rho =r_{0}\varepsilon $ is the screening length and $\varepsilon $
is the dielectric constant of the bulk material. When $r_{0}=1$ and $g=0$
this potential becomes the Coulomb potential, while if $\mathfrak{K}\frac{%
e^{2}}{\rho }=2D$ and $g^{2}=1/2$ it coincides with the Kratzer potential (%
\ref{Kr1}).

\noindent\section{Kratzer potential }

After separation of the center-of-mass and relative motions, for the
relative motion of two interacting particles we obtain the following
equation 
\begin{equation}
\left[ -\frac{\hbar ^{2}}{2\mu }\Delta _{\mathbf{r}}+V(r)\right] \Psi (%
\mathbf{r})=E\Psi (\mathbf{r}),\ \   \label{In3}
\end{equation}%
where $\mu =\frac{m_{1}m_{2}}{m_{1}+m_{2}}$ is the two-particle reduced
mass, $V(r)=V(\left\vert \mathbf{r}_{1}-\mathbf{r}_{2}\right\vert )$, $\Psi (%
\mathbf{r})$ and $E$ are two particles' eigenfunction and eigenenergy,
respectively, and $\Delta _{\mathbf{r}}$ is the Laplace operator in 2D space
with respect to the coordinate of the relative motion $\mathbf{r.}$

Let find the eigenfunctions and eigenenergies of two particles interacting
via the Kratzer potential in 2D configuration space. Due to the central
symmetry of the potential $V(r)$ it is convenient to write the 2D Laplace
operators in the polar coordinates \cite{Morse1953}. Following the standard
procedure \cite{Robinett2006,Zaslow1967} to separate the radial $r$ and
angular variables in 2D space, one can write the radial Schr\"{o}dinger
equation in 2D space with a cental symmetry potential as 
\begin{equation}
\frac{d^{2}\phi (r)}{dr^{2}}+\frac{1}{r}\frac{d\phi (r)}{dr}-\frac{m^{2}}{%
r^{2}}\phi (r)+\frac{2\mu }{\hbar ^{2}}\left[ E-V(r)\right] \phi (r)=0.
\label{In12}
\end{equation}%
In Eq. (\ref{In12}) $m$ is the azimutal quantum number and $\mu $ the
reduced mass of two particles. The 2D radial Schr\"{o}dinger equation (\ref%
{In12}) with potential (\ref{Kr1}) reads 
\begin{equation}
\left( \frac{d^{2}}{dr^{2}}+\frac{1}{r}\frac{d}{dr}-\frac{m^{2}}{r^{2}}%
\right) \phi (r)+\frac{2\mu }{\hslash ^{2}}\left[ E+2D_{0}\left( \frac{r_{0}%
}{r}\ -\frac{1}{2}\ \frac{r_{0}^{2}}{r^{2}}\ \right) \right] \phi (r)=0.\ \ 
\label{Kr2}
\end{equation}%
Let introduce notations 
\begin{equation}
k^{2}=-\frac{2\mu r_{0}^{2}}{\hslash ^{2}}E,\text{ \ }\gamma ^{2}=\frac{2\mu
r_{0}^{2}}{\hslash ^{2}}D_{0}.\   \label{Kr3}
\end{equation}%
For the bound state $k>0$ and $\gamma >0$. Using a new variable $z=\frac{r}{%
r_{0}}$, so $r=r_{0}z$ and notations (\ref{Kr3}) we can rewrite (\ref{Kr2})
in the following form 
\begin{equation}
\frac{d^{2}\phi (z)}{dz^{2}}+\frac{1}{z}\frac{d\phi (z)}{dz}+\frac{%
-k^{2}z^{2}+2\gamma ^{2}z-\gamma ^{2}-m^{2}}{z^{2}}\phi (z)=0.\   \label{Kr7}
\end{equation}%
The differential equation (\ref{Kr7}) has a singularity at $z=0$ and an
irregular singularity when $z\rightarrow \infty $. One can use the
characteristic equation 
\begin{equation}
\alpha ^{2}-m^{2}-\gamma ^{2}=0\ \   \label{Kr71}
\end{equation}%
so 
\begin{equation}
\alpha =\sqrt{m^{2}+\gamma ^{2}}\ \   \label{Kr72}
\end{equation}%
which leads to $\phi (z)\propto z^{\alpha }$ at $z\rightarrow 0,$ when $%
\alpha >0$, and, therefore, the radial wave function vanishes at $z=0$.
Rewriting Eq. (\ref{Kr7}) as 
\begin{equation}
z^{2}\frac{d^{2}\phi (z)}{dz^{2}}+z\frac{d\phi (z)}{dz}+\left( 2\gamma
^{2}z-k^{2}z^{2}-\alpha ^{2}\right) \phi (z)=0\ \   \label{Kr8}
\end{equation}%
and using substitution 
\begin{equation}
\phi (z)=z^{\alpha }e^{-kz}u(z),\   \label{Kr81}
\end{equation}%
after lengthily calculation Eq. (\ref{Kr8}) becomes 
\begin{equation}
z\frac{d^{2}u(z)}{dz^{2}}+(1+2\alpha -2kz)\frac{du(z)}{dz}-\left( k+2k\alpha
-2\gamma ^{2}\right) u(z)=0.\ \   \label{Kr18}
\end{equation}%
Such type equation as (\ref{Kr18}) has an analytical solution in the class
of hypergeometric functions \cite{Polyanin2018}.

After introducing in (\ref{Kr18}) a new variable $x=2kz$ \ this equation can
be written as \cite{Polyanin2018} 
\begin{equation}
x\frac{d^{2}u(x)}{dx^{2}}+(b-x)\frac{du(x)}{dx}-au(x)=0,  \label{Kr20}
\end{equation}%
where $a=\frac{1}{2}+\alpha -\frac{\gamma ^{2}}{k}$ and $b=1+2\alpha $.
Equation (\ref{Kr20}) is the confluent hypergeometric equation, and it is
often called Kummer's equation \cite{Gradshteyn,Abramowitz,Polyanin2018}. \
This equation has two linearly independent solutions \cite{Polyanin2018}. 
\begin{equation}
f(x)=C_{1}\Phi (a,b;x)+C_{2}\Psi (a-b+1,2-b;x)\ \   \label{In63}
\end{equation}%
where $C_{1}$ and $C_{2}$ are the normalization constants. One solution $%
\Phi (a,b;z)$ is regular at the origin $z=0$ and is a confluent
hypergeometric function. The second solution $\Psi (a-b+1,2-b;z)$\ is
diverging. Equation (\ref{Kr20}) is the degenerate hypergeometric equation
which is the general type of Kummer's equation of the confluent
hypergeometric series. Because $b$ is not integer the Kummer series is a
particular solution 
\begin{equation}
u(x)=C_{1}\text{ }_{1}F_{1}\left( \frac{1}{2}+\alpha -\frac{\gamma ^{2}}{k}%
,1+2\alpha ;x\right) .  \label{Kr21}
\end{equation}%
This leads to the radial wave function 
\begin{equation}
\phi (r)=C_{0}\left( \frac{r}{r_{0}}\right) ^{\alpha }e^{-k\frac{r}{r_{0}}%
}\,\,_{1}F_{1}(\frac{1}{2}+\alpha -\frac{\gamma ^{2}}{k},\,\,1+2\alpha
;\,\,2k\frac{r}{r_{0}}),  \label{Kr22}
\end{equation}%
where $C_{0}$ is the normalization constant. The function (\ref{Kr22}) is
normalizable if $\phi (r\rightarrow \infty )\rightarrow 0$. This requires
that 
\begin{equation}
\frac{1}{2}+\alpha -\frac{\gamma ^{2}}{k}=-n,\ n=0,1,2,3,...\   \label{Kr23}
\end{equation}%
Using values for $k$ and $\alpha $ from Eqs. (\ref{Kr3}) and (\ref{Kr72}) we
obtain 
\begin{equation}
E_{nm}=-\frac{\hslash ^{2}}{2\mu r_{0}^{2}}\frac{\gamma ^{4}}{\left( \frac{1%
}{2}+n+\sqrt{m^{2}+\gamma ^{2}}\ \right) ^{2}\ }\ ,\text{ \ }n=0,1,2,3,...%
\text{ \ }m=0,1,2,3,..  \label{Kr24}
\end{equation}%
where $\gamma ^{2}$ is defined in Eq. (\ref{Kr3}). Thus, the energy levels $%
E_{nm}$ are discrete and depend on radial and azimuthal quantum numbers.

A representation of special functions in terms of a confluent hypergeometric
function allows a connection of the confluent hypergeometric function (\ref%
{Kr21}) with the associated Laguerre polynomials $L_{n}^{\alpha}(x)$, $%
\alpha\in R$, \cite{Gradshteyn}

\begin{equation}
L_{n}^{\alpha }(x)=\frac{\Gamma (n+\alpha +1)}{\left( n+1\right) !\Gamma
(\alpha +1)}\text{ }_{1}F_{1}(-n,1+\alpha ;x),
\end{equation}%
where $\Gamma $ is a gamma function. Therefore, the confluent hypergeometric
function is proportional to the associated Laguerre polynomial and the
radial wave function (\ref{Kr22}) can be expressed in terms of $%
L_{n}^{\alpha }(x)$ as: 
\begin{equation}
\phi (r)=C\left( \frac{r}{r_{0}}\right) ^{\alpha }e^{-k\frac{r}{r_{0}}%
}\,\,L_{n}^{2\alpha }\left( 2k\frac{r}{r_{0}}\right) .  \label{Kr255}
\end{equation}%
The normalization constant $C$ can be found through the evaluation of the
integral

\begin{equation}
\int\limits_{0}^{\infty }x^{\alpha +1}e^{-x}\left[ L_{n}^{\alpha }(x)\right]
^{2}dx,\text{ when }\alpha \in R\text{ (integer or real). }
\end{equation}%
To evaluate this integral one can use a formula for recurrence in the index $%
n$ of $L_{n}^{\alpha }(x)$ and orthogonality relation for $L_{n}^{\alpha
}(x) $ \cite{Gradshteyn} 
\begin{equation}
xL_{n}^{\alpha }(x)=\left( 2n+\alpha +1\right) L_{n}^{\alpha }(x)-\left(
n+1\right) L_{n+1}^{\alpha }(x)-(n+\alpha )L_{n-1}^{\alpha }(x),
\label{KrRecur}
\end{equation}

\begin{equation}
\int\limits_{0}^{\infty }x^{\alpha}e^{-x}L_{n}^{\alpha }(x)L_{m}^{\alpha
}(x)dx=\left\{ 
\begin{array}{c}
0,\text{\ \ \ \ \ \ \ \ \ \ \ }n\neq m \\ 
\frac{\Gamma (n+\alpha +1)}{n!},\text{ }n=m%
\end{array}%
\right. .  \label{Krortog}
\end{equation}%
The consideration of (\ref{KrRecur}) and (\ref{Krortog}) leads to the
normalization coefficient of the radial wave function $C=\frac{\left(
2k\right) ^{\alpha +1}}{r_{0}}\sqrt{\frac{n!}{(2n+2\alpha +1)\Gamma
(n+2\alpha +1)}}$. Finally, the eigenfunctions for two particles in
two--dimensional space interacting via the Kratzer potential (\ref{Kr1}) are: 
\begin{equation}
\Psi (r,\varphi )=\frac{\left( 2k\right) ^{\alpha +1}}{r_{0}}\sqrt{\frac{n!}{%
2\pi (2n+2\alpha +1)\Gamma (n+2\alpha +1)}}\left( \frac{r}{r_{0}}\right)
^{\alpha }e^{-k\frac{r}{r_{0}}}\,\,L_{n}^{2\alpha }\left( 2k\frac{r}{r_{0}}%
\right) e^{im\varphi }  \label{Kr26}
\end{equation}

\section{Modified Kratzer's potential 1}

Below we find eigenfunctions and eigenenergies of two particles interacting
via (\ref{Kr31}) potential in 2D configuration space. The radial Schr\"{o}%
dinger equation in 2D space with the potential (\ref{Kr31}) reads%
\begin{equation}
\left( \frac{d^{2}}{dr^{2}}+\frac{1}{r}\frac{d}{dr}-\frac{m^{2}}{r^{2}}%
\right) \phi (r)+\frac{2\mu }{\hslash ^{2}}\left[ E-D_{0}\left( 1-2\frac{%
r_{0}}{r}+\frac{r_{0}^{2}}{r^{2}}\right) \right] \phi (r)=0.  \label{Kr32}
\end{equation}%
Considering notations (\ref{Kr3}) and (\ref{Kr72}) and the same variable $z,$
one can rewrite Eq. (\ref{Kr32}) in the following form 
\begin{equation}
z^{2}\frac{d^{2}\phi (z)}{dz^{2}}+z\frac{d\phi (z)}{dz}+\left[ -\left(
k^{2}+\gamma ^{2}\right) z^{2}+2\gamma ^{2}z-\alpha ^{2}\right] \phi (r)=0.
\label{Kr38}
\end{equation}%
Following the same procedure as for the Kratzer potential (\ref{Kr1}), Eq. (%
\ref{Kr38}) can be simplified using the substitution $\phi (z)=z^{\alpha
}e^{-\beta z}u(z)$ with $\beta =\sqrt{k^{2}+\gamma ^{2}}$. Applying this
substitution, Eq. (\ref{Kr38}) takes the form 
\begin{equation}
z\frac{d^{2}u(z)}{dz^{2}}+\left( 1+2\alpha -2\beta z\right) \frac{du(z)}{dz}-%
\left[ \beta +2\alpha \beta -2\gamma ^{2}\right] u(z)=0.  \label{Kr393}
\end{equation}%
Introducing a new variable $x=2\beta z$\ in Eq. (\ref{Kr393}) leads to%
\begin{equation}
x\frac{d^{2}u(x)}{dx^{2}}+\left( 1+2\alpha -x\right) \frac{du(x)}{dx}-\left[ 
\frac{1}{2}+\alpha -\frac{\gamma ^{2}}{\beta }\right] u(x)=0.
\end{equation}%
The latter equation is the Kummer's equation of the confluent hypergeometric
series, where $1+2\alpha =1+2\left( m^{2}+\gamma ^{2}\right) ^{1/2}$ is not
an integer due to $\gamma ^{2}.$ Hence, (\ref{Kr393}) has a particular
solution 
\begin{equation}
u(2\beta z)\text{ =}_{1}F_{1}(\frac{1}{2}+\alpha -\frac{\gamma ^{2}}{\beta }%
,1+2\alpha ,2\beta z).
\end{equation}%
Finally, the radial wave function in terms of the associated Laguerre
polynomials becomes 
\begin{equation}
\phi (r)=C\left( \frac{r}{r_{0}}\right) ^{\alpha }\text{e}^{-\beta \frac{r}{%
r_{0}}}\text{ }L_{n}^{2\alpha }\left( 2\beta \frac{r}{r_{0}}\right) .
\label{Kr396}
\end{equation}%
The comparison of (\ref{Kr396}) and (\ref{Kr25}) shows that while the radial
wave funcions have the same analytical funtional dependence, they are
different because $k<\beta =\sqrt{k^{2}+\gamma ^{2}}.$ The requirement for
the normalizability of the radial wave function leads to the quantization
condition 
\begin{equation}
\frac{1}{2}+\alpha -\frac{\gamma ^{2}}{\beta }=-n,\ n=0,1,2,3,...\ 
\label{Kr398}
\end{equation}%
and normalization constant $C=\frac{(2\beta )^{\alpha +1}}{r_{0}}\sqrt{\frac{%
n!}{(2n+2\alpha +1)\Gamma (n+2\alpha +1)}}.$ This constant coincides with
the normalization constant for the solution with the Kratzer potential only
when $\beta =k$.

Using values for $\gamma $ from Eq. (\ref{Kr3}), $\alpha =\sqrt{m^{2}+\gamma
^{2}},$ $\beta =\sqrt{k^{2}+\gamma ^{2}}$ in Eq. (\ref{Kr398}) and solving
it for the energy $E$ we obtain 
\begin{equation}
E_{nm}=-\frac{\hslash ^{2}}{2\mu r_{0}^{2}}\frac{\gamma ^{4}}{\left( \frac{1%
}{2}+n+\sqrt{m^{2}+\gamma ^{2}}\ \right) ^{2}\ }\ -D_{0},\text{ \ }%
n=0,1,2,3,...\text{ \ }m=0,1,2,3,..  \label{Kr397}
\end{equation}%
Thus, the energy levels $E_{nm}$ depend on radial and azimuthal quantum
numbers. The comparison of energy spectra (\ref{Kr24}) and (\ref{Kr397})
shows that they have the same dependence on the quantum numbers $n$ and $m$,
but the potential (\ref{Kr31}) leads to the shift of the energy levels by
the dissociation energy $D_{0}$. The total eigenfunction for two particle
interacting via (\ref{Kr31}) potential in terms of the associated Laguerre
polynomials becomes

\begin{equation}
\Psi (r,\varphi )=\frac{(2\beta )^{\alpha +1}}{r_{0}}\sqrt{\frac{n!}{2\pi
(2n+2\alpha +1)\Gamma (n+2\alpha +1)}}\left( \frac{r}{r_{0}}\right) ^{\alpha
}\text{e}^{-\beta \frac{r}{r_{0}}}\text{ }L_{n}^{2\alpha }\left( 2\beta 
\frac{r}{r_{0}}\right) e^{im\varphi }  \label{Kr399}
\end{equation}

\section{Modified Kratzer's potential 2}

By substituting (\ref{Kr25}) into the radial Schr\"{o}dinger equation in 2D
space (\ref{In12}) we obtain 
\begin{equation}
\left( \frac{d^{2}}{dr^{2}}+\frac{1}{r}\frac{d}{dr}-\frac{m^{2}}{r^{2}}%
\right) \phi (r)+\frac{2\mu }{\hslash ^{2}}\left[ E+\frac{e^{2}}{4\pi \rho }%
\left( \frac{r_{0}}{r}\ -g^{2}\frac{r_{0}^{2}}{r^{2}}\ \right) \right] \phi
(r)=0.\ \   \label{MKr1}
\end{equation}%
One can start from (\ref{MKr1}), introduce a variable $r=r_{0}z,$ notation 
\begin{equation}
\text{ \ }\delta ^{2}=\mathfrak{K}\frac{\mu r_{0}^{2}}{\hslash ^{2}}\frac{%
e^{2}}{\rho },\   \label{MKr3}
\end{equation}%
repeat the procedure presented for the Kratzer potential (\ref{Kr1}) using
the substitution $\phi (z)=z^{\eta }e^{-kz}u(z)$ with $\eta $\ $=\sqrt{%
2g^{2}\delta ^{2}+m^{2}}$, and finally obtains 
\begin{equation}
z\frac{d^{2}u(z)}{dz^{2}}+(1+2\eta -2kz)\frac{du(z)}{dz}-(k+2k\eta -2\delta
^{2})u(z)=0.  \label{MKr4}
\end{equation}%
Therefore, (\ref{MKr4}) has an analytical solution that is a confluent
hypergeometric function: $u(z)=\,_{1}F_{1}\left( \frac{1}{2}+\eta -\frac{%
\delta ^{2}}{k},1+2\eta ,2kz\right) $. In terms of \ the associated Laguerre
polynomials the total radial wave function of the exciton bound via (\ref%
{Kr25}) potential reads%
\begin{equation}
u(r)=C\left( \frac{r}{r_{0}}\right) ^{\eta }e^{-k\frac{r}{r_{0}}%
}\,\,L_{n}^{2\eta }\left( 2k\frac{r}{r_{0}}\right) .  \label{MKr5}
\end{equation}%
The radial function (\ref{MKr5}) can be normalized only if $\frac{1}{2}-%
\frac{\delta ^{2}}{k}+\eta =-n;\,\,\,\,\,\,\,\,n=0,1,2,...$ The latter
condition with the values of $k$, $\delta $, and $\eta $ gives the following
quantized energy levels 
\begin{equation}
E_{nm}=-\mathfrak{K}^{2}\frac{e^{4}\mu }{2\hbar ^{2}}\frac{1}{\left( n+\frac{%
1}{2}+\sqrt{2g^{2}\delta ^{2}+m^{2}}\right) ^{2}}.  \label{MKr6}
\end{equation}%
Note that when $g=0$, since $n$ and $m$ are any positive integers included
0, a new quantum number in\ Eq. (\ref{MKr6}) is $n+m=N$. Hence, the energy
levels (\ref{MKr6}) coincide with the spectra for the Coulomb potential \cite%
{Zaslow1967,Yang1991,Portnoi2002,Kezerashvili2023}: 
\begin{equation}
E_{n}=-\mathfrak{K}^{2}\frac{e^{4}\mu }{2\hslash ^{2}}\frac{1}{\left( n+%
\frac{1}{2}\right) ^{2}},\text{ \ \ }n=0,1,2,3,...  \label{19C}
\end{equation}%
While when $g=1/2$ the energy spectrum (\ref{MKr6}) coincides with (\ref%
{Kr24}) if $2D=\mathfrak{K}\frac{e^{2}}{\rho }$. Each energy level (\ref{19C}%
) is ($2n+1$)-fold degenerate, the so-called accidental degeneracy. Notably,
(\ref{19C}) does not contain explicitly the azimuthal quantum number $m$,
which enters the radial equation. From (\ref{19C}) one can conclude that the
spectrum of energy is changes from $E_{n}\sim n^{-2}$ (Rydberg series) in 3D
space \cite{Landau}, to $E_{n}\sim \left( n+1/2\right) ^{-2}$ in 2D space.
Therefore, for example, the ground state energy increases by a factor of 4
in 2D space. Thus, the reduction of dimensionality decreases the kinetic
energy of two particles in 2D space interacting via the Coulomb potential
due to the decrease of the degrees of freedom from three to two.

\begin{figure}[t]
\centering
\includegraphics[width=5.7cm]{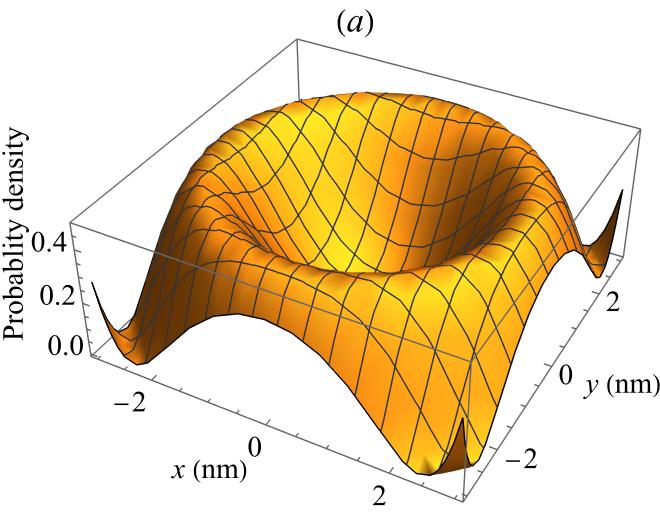} %
\includegraphics[width=5.7cm]{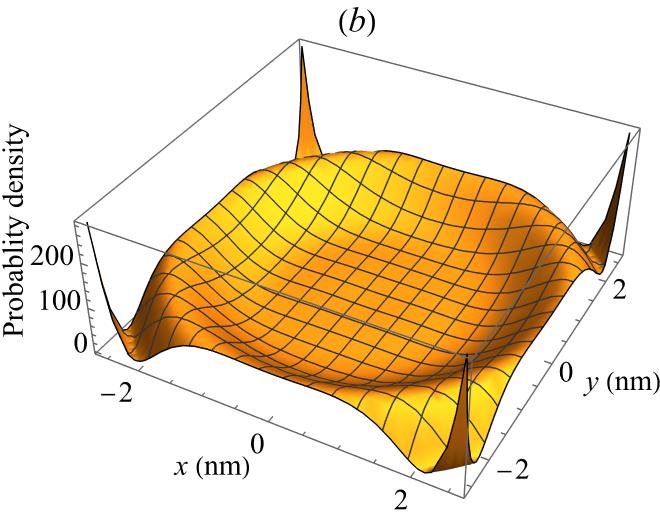} %
\includegraphics[width=5.7cm]{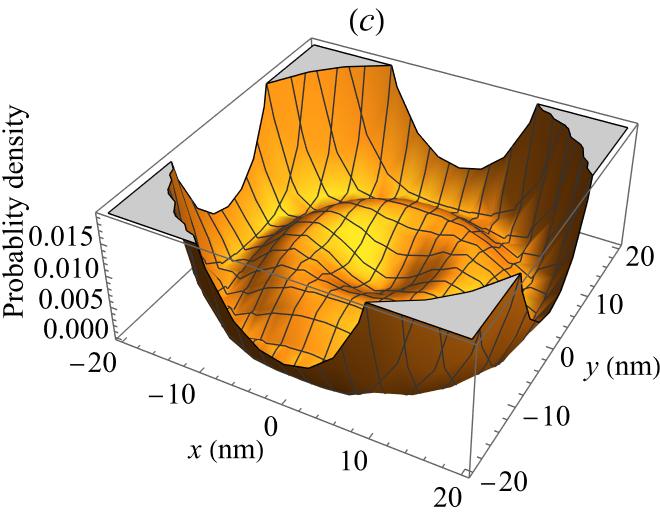}
\caption{(Color online) The probability distribution for two particles
calculated with ($a$) the Kratzer potential (\protect\ref{Kr1}), ($b$) the
modified Kratzer's potential (\protect\ref{Kr31}) for interactions of
diatomic molecules, ($c$) the modified Kratzer's potential (\protect\ref%
{Kr25}) for an electron-hole interaction. Calculations are performed for $%
n=3 $ and $m=1$.}
\label{Fig1}
\end{figure}

The normalization of the radial function (\ref{MKr6}) when $-\frac{1}{2}%
-\eta +\frac{\delta ^{2}}{k}=n$ gives 
\begin{equation}
C=\frac{(2k)^{\eta +1}}{r_{0}}\sqrt{\frac{n!}{(2n+2\alpha +1)\Gamma
(n+2\alpha +1)}}.
\end{equation}%
Hence, the total eigenfunctions related to the eigenenergies (\ref{MKr6}) in
terms of the Laguerre polynomials is 
\begin{equation}
\Psi (r,\varphi )=\frac{(2k)^{\eta +1}}{r_{0}}\sqrt{\frac{n!}{2\pi
(2n+2\alpha +1)\Gamma (n+2\alpha +1)}}\left( \frac{r}{r_{0}}\right) ^{\eta
}e^{-k\frac{r}{r_{0}}}\,\,L_{n}^{2\eta }\left( 2k\frac{r}{r_{0}}\right)
e^{im\varphi }.  \label{MKr8}
\end{equation}%
noindent A careful examination of Eqs. (\ref{Kr26}), (\ref{Kr399}), and (\ref%
{MKr8}) shows that the total eigenfunctions obtained for the Kratzer (\ref%
{Kr1}), modified Kratzer's (\ref{Kr31}), and (\ref{Kr25}) potentials have
the same analytical dependences with different $\alpha $, $\beta $, $\eta $,
and $k$. Nevertheless, the most remarkable feature is that coordinate-space
probability distributions for two particles calculated with the
eigenfuctions (\ref{Kr26}), (\ref{Kr399}), and (\ref{MKr8}) are entirely
different. 
The probability distributions for two particles calculated with the
eigenfuctions (\ref{Kr26}), (\ref{Kr399}), and (\ref{MKr8}) are presented in
Fig. \ref{Fig1}. Obviously, the eigenfunction (\ref{MKr8}) is the same as (%
\ref{Kr26}) when $\eta =\alpha $ and (\ref{MKr8}) coincides with (\ref{Kr399}%
) when $\eta =\alpha $ and $\beta =k$. Under these conditions the
probability distributions for two particles interacting via (\ref{Kr1}), (%
\ref{Kr31}), and (\ref{Kr25}) potentials are the same.

\section{Conclusion}

One of the challenging tasks of quantum mechanics in two-dimension is to
find the exact analytical solution of the Schr\"{o}dinger equation for a
given potential and any arbitrary value $m$ in 2D configuration spaces. This
solution can be further used to define the observables of the system.

In this study, we found the exact bound state solutions of the
two--dimensional Schr\"{o}dinger equation with Kratzer--type potentials and
present analytical expressions for the eigenvalues and eigenfunctions. The
eigenfunctions are given in terms of the associated Laguerre polynomials.

\end{document}